\documentclass[11pt,onecolumn,amssymb,aps,nofootinbib,showpacs]{revtex4-1} %,showpacs
\usepackage{amsmath, amsthm, amscd, amssymb}
\usepackage{bm}
\usepackage{bbm}

\usepackage{feyn}
\let\vec\boldvec%

\begin{document}

\title{The effect of spontaneous collapses on neutrino oscillations}

\author{Sandro Donadi}
\email{sandro.donadi@ts.infn.it} \affiliation{Department of Physics, University of
Trieste, Strada Costiera 11, 34151 Trieste, Italy} \affiliation{Istituto
Nazionale di Fisica Nucleare, Trieste Section, Via Valerio 2, 34127 Trieste,
Italy}

\author{Angelo Bassi}
\email{bassi@ts.infn.it} \affiliation{Department of Physics, University of
Trieste, Strada Costiera 11, 34151 Trieste, Italy} \affiliation{Istituto
Nazionale di Fisica Nucleare, Trieste Section, Via Valerio 2, 34127 Trieste,
Italy}

\author{Catalina Curceanu}
\email{Catalina.Petrascu@lnf.infn.it} \affiliation{INFN Laboratori Nazionali di Frascati, Via E. Fermi 40, 99944 Frascati
(Roma), Italy}

\author{Luca Ferialdi}
\email{ferialdi@ts.infn.it} \affiliation{Department of Physics, University of
Trieste, Strada Costiera 11, 34151 Trieste, Italy} \affiliation{Istituto
Nazionale di Fisica Nucleare, Trieste Section, Via Valerio 2, 34127 Trieste,
Italy}

\begin{abstract}
We compute the effect of collapse models on neutrino oscillations. The effect of the collapse is to modify the evolution of the {\it spatial} part of the wave function and we will show that this indirectly amounts to a change on the flavor components. For the analysis we use the mass proportional CSL model, and perform the calculation to second order perturbation theory.  As we will show, the CSL effect is very small---mainly due to the very small mass of neutrinos---and practically undetectable.
\end{abstract}
\maketitle

\section{Introduction}

The general validity of the superposition principle of quantum mechanics is questioned by an increasing number of scientists~\cite{Bel,Ad,We,Le}, and is subject to an intense experimental verification~\cite{Ar1,Ar2,Ar3,Ar4,Bo,Mo}. The possibility that the superposition principle might have only a limited range of validity is foreseen by collapse models~\cite{Grw,Cslmass,Fu,Csl,adlerphoto,Pr,Ad2,Pe1,Pe2,Pe3,Di1,Di2}, which predict small deviations from standard quantum mechanics, for all those cases where quantum linearity plays a fundamental role. Neutrino oscillations~\cite{Beuthe,Neu1,Neu2,Neu3} are one such a case, and the goal of this article is to present the theoretical analysis and estimate of the effect of spontaneous collapses on the oscillatory behavior of neutrinos. 

A previous analysis of this kind was proposed in~\cite{Ch}, based on the Penrose model of gravity induced collapse~\cite{Pd1,Pd2}. This model however lacks a fully consistent dynamical equation, and previous attempts to fill this gap~\cite{Dg} have been criticized~\cite{Gpd}. Moreover, as shown in~\cite{Gpd}, the model fails when applied to single constituents (protons, electrons, ...), since in this case its predictions are in conflict with known experimental data. Therefore, the application of gravity induced collapse models to neutrino oscillations is rather delicate. 

Here we will compute the spontaneous collapse effect on neutrino oscillations using the mass proportional version~\cite{Cslmass,Fu} of the CLS model~\cite{Csl}, which is widely used in physical applications, together with the GRW model~\cite{Grw}. Its dynamics is described by the following stochastic differential equation:
\begin{equation} \label{eq:csl-massa}
d|\phi_{t}\rangle=\left[-\frac{i}{\hbar}Hdt+\frac{\sqrt{\gamma}}{m_{0}}\int d\mathbf{x}\,\left(M(\mathbf{x})-\left\langle M(\mathbf{x})\right\rangle \right)dW_{t}(\mathbf{x})-\frac{\gamma}{2m_{0}^{2}}\int d\mathbf{x}\,\left(M(\mathbf{x})-\left\langle M(\mathbf{x})\right\rangle \right)^{2}dt\right]|\phi_{t}\rangle,
\end{equation}
where the operator $H$ is the standard quantum Hamiltonian of the system and the other two
terms induce the collapse of the wave function in space. The mass $m_0$ is a reference mass, which is taken equal to that of a nucleon. The parameter $\gamma$ is a positive coupling constant which sets the strength of the collapse process and $\left\langle M(\mathbf{x})\right\rangle =\left\langle \phi_{t}\left|M(\mathbf{x})\right|\phi_{t}\right\rangle $, where $M({\bf x})$ is a smeared mass density operator:
\begin{equation}
M\left(\mathbf{x}\right)=\underset{j}{\sum}m_{j}\underset{s}{\sum}\int
d\mathbf{y}g\left(\mathbf{x-y}\right)
\psi_{j}^{\dagger}\left(\mathbf{y},s\right)\psi_{j}\left(\mathbf{y},s\right),
\end{equation}
$\psi_{j}^{\dagger}\left(\mathbf{y},s\right)$,
$\psi_{j}\left(\mathbf{y},s\right)$ being, respectively, the creator and
annihilation operators of a particle of type $j$, having mass $m_j$ and spin $s$, in the space point
$\mathbf{y}$. The smearing function $g({\bf x})$ is taken equal to
\begin{equation} \label{eq:nnbnm}
g(\mathbf{x}) \; = \; \frac{1}{\left(\sqrt{2\pi}r_{C}\right)^{3}}\;
e^{-\mathbf{x}^{2}/2r_{C}^{2}},
\end{equation}
where $r_C$ is the other new phenomenological constant of the model. Finally, $W_{t}\left(\mathbf{x}\right)$ is an
ensemble of independent standard Wiener processes, one for each point in space. The standard numerical value of the correlation length $r_C$ is~\cite{Csl}:
\begin{equation}
r_C \; \simeq \; 10^{-5}\text{cm},
\end{equation}
while, in the literature, two different values for the collapse strength $\gamma$ have been proposed. The first value has been originally proposed by Ghirardi, Pearle and Rimini~\cite{Csl}:
\begin{equation}
\gamma \; \simeq \; 10^{-30}\text{cm}^{3}\text{s}^{-1}
\end{equation}
in analogy with the GRW model~\cite{Grw}. The second value has been proposed by Adler, inspired by the analysis of the process of latent image formation according to collapse models, and amounts to~\cite{adlerphoto}:
\begin{equation}
\gamma \; \simeq \; 10^{-22}\text{cm}^{3}\text{s}^{-1}.
\end{equation}

Aim of this work is to understand if, as claimed in~\cite{Ch}, neutrino oscillations can be used to test collapse models, and in particular to improve the upper bounds on the collapse strength $\gamma$. The idea is the following: it is well known that, since flavour eigenstates are linear superposition of mass eigenstates, standard quantum mechanics predicts neutrino oscillations. In the CSL model the dynamics is driven by Eq.~\eqref{eq:csl-massa}, which differs from the Schr\"odinger equation for the two terms accounting for the collapse in space of the wave function. As a consequence, the CSL model predicts a different time evolution of mass eigenstates with respect to that of standard quantum mechanics. This implies, as an indirect consequence, that also flavour eigenstates evolve differently, and therefore that  neutrinos are expected to oscillate in a different manner. In some sense, it is as if neutrinos were traveling through a random medium, instead of free space. It is well known that neutrino oscillations are affected by a random medium~\cite{Sm1,Sm2}. However, we stress that this is more a mathematical analogy as in our case the origin of the randomness is different, and is due to the spontaneous collapse of the wave function. The fact that the collapse mechanism acts on the spatial part of the wave function implies that we have to consider the whole Hilbert space of the system, not just the part related to the flavour degrees of freedom. Technical details about this issue are given in Sec.V.

The paper is organized as follows: in Section II we state the main result of the computation and quantify the damping of neutrino oscillation, as predicted by the mass proportional CSL model. In Section III we compare our result with that of~\cite{Ch}. In Section IV we discuss decoherence effects on neutrino oscillation and compare them with the collapse effects. The remaining sections of the paper are devoted to computing the formulas, which are used in Section II.

\section{The CSL prediction for neutrino oscillation}

According to the mass proportional CSL model, the transition probability of finding a neutrino in a flavour eigenstate $\beta$, when it was initially in the flavour state $\alpha$, is:
\begin{equation} \label{eq:dfggkdse}
P_{\alpha\rightarrow\beta}
=
\sum_{k=1}^{n}\text{U}_{\alpha k}\text{U}_{\beta k}\text{U}_{\alpha k}\text{U}_{\beta k}+\sum_{{k \neq j}}^{n}\text{U}_{\alpha k}\text{U}_{\beta k}\text{U}_{\alpha j}\text{U}_{\beta j} \; e^{-\xi_{jk}t} \; \cos\left[\frac{1}{\hbar}(E_{i}^{\left(k\right)}-E_{i}^{\left(j\right)})t\right].
\end{equation}
$\text{U}$ is the $n\times n$ mixing matrix, which relates the flavour and mass bases, $E_i^{(k)}$ is the initial energy of the neutrino in the mass eigenstate with mass $m_k$, and:  
\begin{equation}\label{eq:xi}
\xi_{jk} \; \equiv \;  \frac{\gamma}{16\pi^{3/2}r_{C}^{3}m_{0}^{2}c^{4}}\left(\frac{m_{j}^{2}c^{4}}{E_{i}^{\left(j\right)}}-\frac{m_{k}^{2}c^{4}}{E_{i}^{\left(k\right)}}\right)^{2}
\end{equation}
is the decay-rate of neutrino oscillations, as predicted by collapse models. Note that the frequency of the oscillations is the same as the one predicted by quantum mechanics\footnote{Eq.~\eqref{eq:dfggkdse} differs from those one typically finds in the literature only because the latter are written in the ultra-relativistic approximation $E=\sqrt{\mathbf{p}^{2}c^{2}+m^{2}c^{4}}\simeq pc\left(1+\frac{m^{2}c^{4}}{2p^{2}c^{2}}\right)$.}. The prediction of the CSL model differs from the standard formula only for a damping factor in front of the oscillating term, with a decay rate given by Eq.~\eqref{eq:xi}. Eqs.~\eqref{eq:dfggkdse} and~\eqref{eq:xi}, which will be derived in Sections V-VIII are significant because they allow to precisely quantify the collapse effect on neutrino oscillations. Since collapse master equations have the same structure as decoherence master equations for open quantum systems, it does not come as a surprise that Eq.~\eqref{eq:dfggkdse} is in agreement with general arguments, which fix the form that the damping terms coming from decoherence effect should take~\cite{floreanini,deco1,deco2}. 

Having the above equations at hand, we can give a quantitative estimate the CSL effect on neutrino oscillations, by computing the damping factor in Eq.~\eqref{eq:dfggkdse}. We consider the stronger value $\gamma\simeq \; 10^{-22}\text{cm}^{3}\text{s}^{-1}$ for the collapse parameter, suggested in~\cite{adlerphoto}.  Substituting the numeric values of the constants in Eq.~\eqref{eq:xi}, one finds that:
\begin{equation}\label{eq:xi2}
\xi_{ij}t\simeq (7.33 \times 10^{-36} \text{s$^{-1}$eV$^2$}) \frac{t}{E^2} \,.
\end{equation}
Here we have taken the largest possible squared mass difference $m_1^2c^4-m_2^2c^4 = 7.59 \times 10^{-5} \text{eV$^2$}$~\cite{deltamass}, where $m_1$ and $m_2$ are respectively the first and the second mass eigenstate and we have considered the ultra-relativistic approximation $E_{i}^{\left(j\right)}=\sqrt{\mathbf{p}_i^{2}c^{2}+m_j^{2}c^{4}}\simeq p_ic\equiv E$, and the same approximation for $E_{i}^{\left(k\right)}$. The energy $E$ and the time $t$ are free and depend on the nature of the neutrinos under study. Neutrinos detected in laboratories have mainly three origins: cosmogenic, solar and those produced in labs. Table I displays the typical values for the energy and time of flight (for simplicity, we assume that neutrinos travel at the speed of light) for these three types of neutrinos. The magnitude of the damping of the oscillations, as predicted by the mass proportional CSL model, has been evaluated using Eq.~\eqref{eq:xi2}. As we can see, in all three cases the CSL damping effect on neutrino oscillations is very small. The main reason is that the masses here involved---those of neutrinos---are very small, thus hampering the collapse mechanism. 

To conclude we compare the effect of the CSL model for the oscillation formula with the error embodied in the ultra-relativistic approximation. This approximation is usually done in the literature since the error introduced is very small. In order to estimate the error due to the ultra-relativistic approximation, we expand in series the energies $E_i$ of Eq.~\eqref{eq:xi} in the ultra-relativistic regime ($pc>>mc^{2}$). The energy difference, at the second order, becomes:
\begin{equation}
E_{i}^{\left(k\right)}-E_{i}^{\left(j\right)}\simeq\frac{m_{k}^{2}c^{4}-m_{j}^{2}c^{4}}{2p_{i}c}-\frac{1}{8}\left(\frac{m_{k}^{4}c^{8}-m_{j}^{4}c^{8}}{p_{i}^{3}c^{3}}\right)\simeq\frac{\triangle m^{2}}{2E}-\frac{\triangle m^{2}}{8E^{3}}\left(m_{k}^{2}c^{4}+m_{j}^{2}c^{4}\right)
\end{equation}
where we introduced $\triangle m^{2}=m_{k}^{2}c^{4}-m_{j}^{2}c^{4}$
and we approximated, only for the denominator, $E_{i}^{\left(k\right)}\simeq E_{i}^{\left(j\right)}\simeq p_{i}c:=E$.
The first term is the oscillation frequency usually considered in the literature~\cite{deco1}. The other
one is the most relevant correction. Even taking the upper
value for neutrinos masses of order of 2.2 eV~\cite{upper} and considering the case of solar neutrinos, those having a
lower energy ($E=10^{6}$ eV), the second term on the right hand side
of the above equation is 12 order of magnitude smaller than the first
term. This shows that the ultra-relativistic approximation is very good in general.

The error in the oscillation formula due to the ultra-relativistic approximation is:
\begin{equation}
\frac{\triangle m^{2}}{8E^{3}}\left(m_{k}^{2}c^{4}+m_{j}^{2}c^{4}\right)\frac{t}{\hbar}=\left(1.01\times10^{10}\;\textrm{s}^{-1}\textrm{eV}^{3}\right)\frac{t}{E^{3}}.
\end{equation}
Using the data in table I, the effect is $\sim10^{-29}$ for cosmogenic neutrinos, $\sim10^{-6}$ for solar neutrinos and $\sim10^{-22}$ for laboratory neutrinos. Compared with the effect due to the collapse, reported in the last line of table I, the error due to the ultra-relativistic approximation is bigger. This is the reason why we did not make such an approximation.
\begin{center}
\begin{table}
  \begin{tabular}{| c| c| c |c| }
    \hline
   & cosmogenic & \hspace{0.5cm}solar\hspace{0.5cm} & laboratory \\ \hline
    E(eV)& $10^{19}$& $10^{6}$ & $10^{10}$ \\ \hline
     t(s)& $3.15\times10^{18}$ & $5\times10^2$ & $2,13\times10^{-2}$\\ \hline\hline
     $\xi_{ij} t$ & $2.31\times10^{-55}$ & $3.66\times10^{-45}$ &$1.56\times10^{-57}$\\
     \hline
  \end{tabular}
\caption{We consider three types of neutrinos: cosmogenic, solar and laboratory neutrinos. For each type, the table shows: the typical order of magnitude of the energies (first line), the time of flight (second line) and the damping factor as predicted by the mass proportional CSL model (third line).}
\end{table}
\end{center}

\section{Comparison with the Diosi-Penrose gravity-induced collapse model}

As mentioned in the introduction, the damping of neutrino oscillations due to gravitational collapse, as described by the Diosi-Penrose model, was first studied in~\cite{Ch}. In this work, the author argued that the decaying factor (the analog of $\xi_{jk} t$ for the CSL model) has the following form:
\begin{equation}
\Lambda_{G}^{j,k}\equiv \int_{D}^{L}\triangle E_{G}^{j,k}\left(L'\right)dL'\,,
\end{equation}
where $D$ is the distance such that $\triangle E_{G}^{j,k}\left(D\right)=0$, $L$ is the distance traveled by the neutrino, and
\begin{equation}
\triangle E_{G}^{j,k}=4\pi\bar{\xi}\int\int\frac{\left[\rho_{j}\left(\mathbf{r}\right)-\rho_{k}\left(\mathbf{r}\right)\right]\left[\rho_{j}\left(\mathbf{r}'\right)-\rho_{k}\left(\mathbf{r}'\right)\right]}{\left|\mathbf{r}-\mathbf{r}'\right|}d\mathbf{r}d\mathbf{r}'\,,
\end{equation}
where $\bar{\xi}$ is a parameter that we will set equal to $-G$, with $G$ the gravitational constant, like in the original paper by Penrose~\cite{Pd1}. Moreover, $\rho_{1}\left(\mathbf{r}\right)$ and $\rho_{2}\left(\mathbf{r}\right)$ are the two mass distributions, one for each different mass eigenstate. Since these two distributions travel at different velocities, $\triangle E_{G}^{j,k}$ has a dependence on the traveled distance $L'$ (see Eq.~(14) of~\cite{Ch}). Following the computation done in~\cite{Ch} and keeping all constants explicit, one finds: 
\begin{equation}
\Lambda_{G}^{j,k}\simeq8\pi \frac{G}{\hbar c}\left[\frac{3\left(m_{j}+m_{k}\right)\hbar^{2}}{5G_{F}}-\frac{m_{j}m_{k}E}{2\pi\hbar c}\ln\left(\frac{6\left(m_{j}+m_{k}\right)\pi\hbar^{3}c}{5m_{j}m_{k}G_{F}E}\right)\right]L
\end{equation}
where $G_{F}$ is the Fermi constant and $m_{j}, m_k$ are the neutrino masses.
In the case of cosmogenic neutrinos which have an energy of about $E=10^{19}$eV and travel a distance $L\simeq10^{25}$m, the magnitude of the damping factor $\Lambda_G^{j,k}$ lies between 1 and $10^{-2}$,
depending on the mass of the neutrino. Therefore, for the Diosi-Penrose model damping effect is by far stronger than that computed with the CSL model. This does not come as a surprise, because---as we stated in the introduction---it is well known that the model predicts, for single elementary constituents, a too-strong collapse of the wave function, which is incompatible with known experimental data~\cite{Gpd}. The reason is the following. It is clear that the model gives rise to divergences in the point-like limit, therefore one has to introduce a cutoff~\cite{Dg2}. One way of doing it, is to consider elementary particles as spherical mass-distributions with a finite radius $R$. In~\cite{Dg} it was proposed to take $R \sim 10^{-15}$m, i.e. the nuclear size. However in~\cite{Gpd} it was shown that also in this case the model is consistent with known facts (the energy increase of isolated systems, due to the collapse, is too large), and proposed a much larger radius, namely $R \sim 10^{-5}$cm, in order to restore compatibility. On the contrary, in~\cite{Ch} the radius ($a_j$, according to the paper's notation) $R \sim G_F m / \hbar^2 \sim 10^{-31\pm1}$m (where the uncertainty depends on the chosen value for the neutrino's mass)  was considered. This cutoff is too small, therefore the result cannot be trusted.

\section{Decoherence effects}

While traveling through the Universe and in particular through the atmosphere, neutrinos interact with the surrounding environment and scatter with other particles, mainly protons, electrons and other neutrinos. These interactions give rise to decoherence effects which also modify  neutrino oscillations. Since protons interact with neutrinos only via the neutral weak current, both neutrino families\footnote{In the following, we will consider decoherence effects only on electronic and muonic neutrinos.} are affected in the same way by this kind of interaction and neutrino oscillations are not modified.
Unlike protons, electrons and neutrinos interact with the incoming neutrino both via neutral and charged weak currents: since these scatterings have different cross sections depending on the neutrino flavor, they contribute to decoherence~\cite{Stol2}.

A natural phenomenological estimate for the order of the decoherence rate is: $\Lambda_{\text{\tiny DEC}}\sim n\,v\,\sigma$ with $v$ the  relative velocity of the incoming neutrinos, $n$ the density of the environmental leptons, and $\sigma$ the relevant scattering cross section whose values are known in the literature~\cite{sigmanue,sigmanunu}:
\begin{eqnarray}
\label{eq:s1}\sigma_{\nu_{e},e}&\simeq&7\times10^{-42}(\mathrm{E}_{\nu}/\mathrm{GeV})\mathrm{cm}^2\,,\\
\sigma_{\nu_{\mu},e}&\simeq&10^{-42}(\mathrm{E}_{\nu}/\mathrm{GeV})\mathrm{cm}^2\,,\\
\sigma_{\nu_e,\nu_e}&\simeq&2,8\times10^{-47}(\mathrm{E}_{\nu}/\mathrm{GeV})\mathrm{cm}^2\,,\\
\label{eq:s2}\sigma_{\nu_e,\nu_{\mu}}&\simeq&4\times10^{-48}(\mathrm{E}_{\nu}/\mathrm{GeV})\mathrm{cm}^2\,.
\end{eqnarray}
The average density of electrons in the outer space and in the atmosphere are respectively $n_e^{\text{\tiny OUT}}\sim1/\mathrm{m}^3$ and
$n_e^{\text{\tiny ATM}}\sim2\times10^{26}/\mathrm{m}^3$,
while the average density of neutrinos is about $n_{\nu}\sim10^8/\mathrm{m}^3$ everywhere.
Assuming $v$ to be the velocity of light in vacuum, then one finds:
\begin{eqnarray}
\Lambda_{\text{\tiny DEC}}^{\text{\tiny OUT}} \sim 10^{-43}
(E/\text{eV})\,\text{Hz}\,,\qquad
\Lambda_{\text{\tiny DEC}}^{\text{\tiny ATM}}\sim10^{-20}(E/\text{eV})
\,\text{Hz}.
\end{eqnarray}
with $\Lambda_{\text{\tiny DEC}}^{\text{\tiny OUT[ATM]}}$ the decoherence rate in the out-space [atmosphere]. Neutrinos travel through the atmosphere within $\sim10^{-4}\,$s. 
Using this data with the time-of-flights and energies in Table~I, for the decoherence damping factor of the cosmogenic neutrinos, one finds: $\sim10^{-5}$. For solar neutrinos instead, one gets: $\sim10^{-18}$,
thus the damping of solar neutrino oscillations is hardly detectable, in agreement with experimental results~\cite{decoexp,vogel}.

This estimate shows that, since environmental decoherence on neutrino oscillations is much stronger than the CSL collapse effect, these spontaneous collapse effects cannot be observed experimentally, even if the technology were sophisticated enough to reach such sensitivities. They would anyhow be masked by decoherence effects. Moreover, decoherence effects are not far away from the collapse effect predicted in~\cite{Ch}, which---as we argued---is anyhow overestimated. Therefore, also gravity-induced collapse effects cannot be detected via neutrino oscillations.

The rest of the paper is devoted to deriving Eqs.~\eqref{eq:dfggkdse} and~\eqref{eq:xi}. The calculation is lengthy but instructive because it shows, as already stressed in the introduction, how the collapse, which acts on the spatial part of the wave function, as a byproduct also affects the flavour degrees of freedom. Moreover, a precise calculation clears any possible misunderstanding about the effect of collapse models on neutrino oscillations.

\section{Mathematical setup}

Working with non-linear equations such as the CSL equation is notoriously difficult. As shown e.g. in~\cite{Im}, the experimentally testable predictions of the model---when averaged over the noise---do not change if the real noise $W_{t}\left(\mathbf{x}\right)$ is replaced by an imaginary noise
$i W_{t}\left(\mathbf{x}\right)$. In this way, one loses the collapse properties of the equation. However, the advantage of having an imaginary noise is that the evolution is described by a standard linear Schr\"odinger equation with a random Hamiltonian:
\begin{equation} \label{eq:htot}
H_{\text{\tiny TOT}} = H - \hbar \sqrt{\gamma} \sum_j \frac{m_{j}}{m_{0}}
\sum_s\int d\mathbf{y}\, w(\mathbf{y},t)
\psi_{j}^{\dagger}(\mathbf{y},s) \psi_{j}(\mathbf{y},s),
\end{equation}
where
\begin{equation} \label{eq:sfsoi}
w(\mathbf{y},t) = \int d\mathbf{x}\, g(\mathbf{x-y})\xi_{t}(\mathbf{x}),
\end{equation}
and $\xi_{t}(\mathbf{x}) = dW_{t}(\mathbf{x})/dt$ is a white noise field, with mean equal to zero and
correlation function $\mathbb{E}[ \xi_{t}(\mathbf{x}) \xi_{s}(\mathbf{y})] = \delta(t-\nobreak s) \delta({\bf x-y})$. As such, $w(\mathbf{x},t)$ is a Gaussian noises field, with zero mean and correlation function:
\begin{equation} \label{eq:sdfddas}
{\mathbb E}[w(\mathbf{x},t) w(\mathbf{y},s)] \; = \;
\delta(t-s)F({\bf x} - {\bf y}), \qquad  F({\bf x}) \; = \;
\frac{1}{(\sqrt{4 \pi} r_C)^3} e^{-{\bf x}^2/4 r_C^2}.
\end{equation}

We are interested in the relativistic generalization of the Hamiltonian~(\ref{eq:htot}). The most natural choice is: $H_{\text{\tiny TOT}} = H_{\text{\tiny D}} + N(t)$ where (in the case of just one type of particle):
\begin{equation}
H_{\text{\tiny D}}=\int d\mathbf{x}\mathcal{H}_{\text{\tiny D}}\left(x\right)=\int d\mathbf{x}\, \psi^{\dagger}\left(x\right)\left[-i\hbar c\,{\vec{\alpha}}\cdot\vec{\nabla}+mc^{2}{\beta}\right]\psi\left(x\right)
\end{equation}
is the standard Dirac Hamiltonian. Here we have introduced the four-vector notation $x \equiv (ct, {\bf x})$, $\psi\left(x\right)$ is the Dirac spinor field, $c$ the speed of light, $m$ the mass of the particle associated to this field, $\vec{\alpha} \equiv (\alpha^1, \alpha^2, \alpha^3)$ with $\alpha^{i}=\gamma^{0}\gamma^{i}$ and $\beta=\gamma^{0}$ where the $\gamma^{\mu}$ are the Dirac matrices, which we take in their standard representation:
\begin{equation}
\gamma^{0}=\left(\begin{array}{cc}
1 & 0\\
0 & -1\end{array}\right),\qquad\gamma^{i}=\left(\begin{array}{cc}
0 & \sigma^{i}\\
-\sigma^{i} & 0\end{array}\right)\qquad\textrm{with}\;\;\; i=1,2,3.
\end{equation}
where $1$ is the identity matrix in two dimensions and $\sigma^{i}$ are the Pauli matrices. The noise term instead is given by:
\begin{equation} \label{eq:dfgdf}
N\left(t\right)=\int d\mathbf{x}\,\mathcal{N}\left(x\right)=-\hbar\sqrt{\gamma_{m}}\int d\mathbf{x}w\left(x\right)\overline{\psi}\left(x\right)\psi\left(x\right), \qquad \gamma_{m}=\gamma\left(\frac{m}{m_{0}}\right)^{2}
\end{equation}
It can be shown\footnote{A proof of this can be found in section 2.3 of~\cite{Greiner}. There, the proof is worked out for the Dirac equation coupled with an electromagnetic field. This is the same case as ours, if one sets $\mathbf{A}\left(x\right)=0$ and $eA_{0}\left(x\right)=-\hbar\sqrt{\gamma_{m}}w\left(x\right)$.} that the Hamiltonian $H_{\text{\tiny TOT}} = H_{\text{\tiny D}} + N(t)$ defined in this way, in the non relativistic limit, reduces to the Hamiltonian of Eq.~(\ref{eq:htot}).

We chose to treat neutrinos as Dirac particles, despite the fact that is not yet known if they are Dirac or Majorana particles. We expect that the size of the collapse effect does not change significantly if Majorana fields are used in place of Dirac fields. As a matter of fact, in~\cite{Kaons} the effect of the CSL model on kaon oscillations formula was studied. Despite the fact that kaons are different from neutrinos and that in~\cite{Kaons} they were studied in the non-relativistic regime, the result is the same as that of this paper: a damping factor in front of the oscillating term, with a decay rate equivalent of to Eq.~\eqref{eq:xi} when the non-relativistic limit is taken.

Working with plane waves gives rise to unphysical infinities, since they are not normalizable. To avoid potential problems, we use the box normalization, i.e. we confine our fields in a box of length $L$, and we impose periodic boundary conditions: $\psi\left(t,\mathbf{x}\right)=\psi\left(t,\mathbf{x}+\mathbf{L}\right)$, where $\mathbf{L}$ is a vector with all the components equal to $L$. In turn, the momentum is discretized: ${\bf p}=\frac{2\pi\hbar}{L}{\bf k}$ with $k^{i}\in\mathbb{Z}$ and $i=1,2,3$ labels the spatial components. Then the Dirac field, in the interaction picture where we choose $H_{\text{\tiny D}}$ as the unperturbed Hamiltonian and $N\left(t\right)$ as the perturbation (the noise coupling $\sqrt{\gamma_{m}}$ is very small), takes the usual expression\footnote{Here and in the following we use the notation of~\cite{Greiner}.}:
\begin{equation} \label{eq:wave}
\psi_{I}\left(x\right)=\sum_{s=1}^{2}\sum_{\mathbf{k}=-\infty}^{+\infty}\frac{1}{\sqrt{L^{3}}}\sqrt{\frac{mc^{2}}{E_{p}}}\left[b\left(p,s\right)u\left(p,s\right)e^{-\frac{i}{\hbar}E_{p}t+i\frac{2\pi}{L}\mathbf{k}\cdot\mathbf{x}}+d^{\dagger}\left(p,s\right)v\left(p,s\right)e^{\frac{i}{\hbar}E_{p}t-i\frac{2\pi}{L}\mathbf{k}\cdot\mathbf{x}}\right],
\end{equation}
where $u$ and $v$ are the usual Dirac spinors, $E_{p}=\sqrt{\mathbf{p}^{2}c^{2}+m^{2}c^{4}}$ is the energy and $b$ and $d$ are operators satisfying the standard anti-commutation relations.
We also recall the relation between the evolution operator $U(t)$ in the Schr\"odinger picture and $U_{I}\left(t\right)$ in the interaction picture~\cite{Sakurai}:
\begin{equation}\label{evolution}
U\left(t\right)=e^{-\frac{i}{\hbar}H_{\text{\tiny D}}t}U_{I}\left(t\right),
\end{equation}
and we set the initial time to 0.

One can question why we use a relativistic Hamiltonian in a model which is not relativistic, since the correlation function of the noise is not Lorentz-invariant. Our approach to this issue, is that collapse models are phenomenological models emerging from a pre-quantum theory yet to be discovered. The noise field is a real cosmological field (whose nature is yet to be investigated) which naturally defines a privileged frame, most likely corresponding to the co-moving frame of the universe. Hence we see no contradiction in analyzing relativistic phenomena with the CSL model. For attempts towards a fully relativistic formulation of collapse models, one can refer to~\cite{Tu,Bedingham}.

\section{The transition amplitude}

In the more general approach to the problem of neutrino oscillations, we consider $n$ flavour eigenstates, which will be labeled by greek subscripts $\left|\nu_{\alpha}\right\rangle$  and each of them is a linear combination of $n$ mass eigenstates that will be labeled by latin subscripts $\left|\nu_{j}\right\rangle$:
\begin{equation}
\left|\nu_{\alpha}\right\rangle =\sum_{j=1}^{n}\text{U}_{\alpha j}\left|\nu_{j}\right\rangle.
\end{equation}
Here, $\text{U}$ is the $n\times n$ mixing matrix, which relates the two different bases; since the flavour eigenstates are supposed to be orthonormal, as well as the mass eigenstates, $\text{U}$ must be unitary. 

We take a neutrino in an initial flavour eigenstate, and compute the probability of finding it in another flavour eigenstate, after some time $t$, assuming that the dynamics is governed by the Hamiltonian~$H_{\text{\tiny TOT}}$. We assume that the neutrino has definite initial and final momenta. This means that its initial and final states are plane waves and that both mass eigenstates have the same momenta. As discussed in ~\cite{Beuthe}, in order to have a more consistent description, one should use wave packets instead of plane waves. However, a wave packet analysis goes beyond the scope of this paper: it would make the calculations much more difficult, and the expected result should not be much different from the one here derived. Moreover, in the standard treatment of neutrino oscillations, the plane wave analysis already gives a satisfactory description, to some degree, both in vacuum and in matter. Mathematically, we will compute the following quantity:
\begin{equation}\label{eq:Tab}
T_{\alpha\rightarrow\beta}\equiv\left\langle \nu_{\beta};\mathbf{p}_{f},s_{f}\left|U\left(t\right)\right|\nu_{\alpha};\mathbf{p}_{i},s_{i}\right\rangle =\sum_{i,j=1}^{n}\text{U}_{\alpha j}\text{U}_{\beta i}^{*}\left\langle \nu_{i};\mathbf{p}_{f},s_{f}\left|U\left(t\right)\right|\nu_{j};\mathbf{p}_{i},s_{i}\right\rangle 
\end{equation}
here $U\left(t\right)$ is the time-evolution operator of Eq.~\eqref{evolution} while $\left|\nu_{\alpha};\mathbf{p}_{i},s_{i}\right\rangle$  is the flavour eigenstate $\alpha$ with initial momentum $\mathbf{p}_{i}$ and spin $s_{i}$ and $\left|\nu_{\beta};\mathbf{p}_{f},s_{f}\right\rangle$  is the final state with momentum $\mathbf{p}_{f}$ and spin $s_{f}$.
Since the Hamiltonian is the sum of Hamiltonians associated to different mass eigenstates ($H=\sum_{j=1}^{n}H_{j}$), it is convenient to expand the flavour eigenstates into the mass eigenstates, as we did in Eq.~(\ref{eq:Tab}). The form of the Hamiltonian also implies that $U(t)$ factorizes: $U\left(t\right)=\prod_{k=1}^{n}U_{k}\left(t\right)$. Here $U_{k}\left(t\right)$ is the time evolution operator related to the Fock space of the neutrino having a definite mass $m_{k}$. This is an important property, because it implies that if $i\neq j$:
\begin{eqnarray}
\langle \nu_{i};\mathbf{p}_{f},s_{f}|U(t)|\nu_{j};\mathbf{p}_{i},s_{i}\rangle  & = & \left\langle \Omega_{1}\left|U_{1}\left(t\right)\right|\Omega_{1}\right\rangle ...\left\langle \nu_{i};\mathbf{p}_{f},s_{f}\left|U_{i}\left(t\right)\right|\Omega_{i}\right\rangle ...\left\langle \Omega_{j}\left|U_{j}\left(t\right)\right|\nu_{j};\mathbf{p}_{i},s_{i}\right\rangle... \nonumber \\
& & ...\left\langle \Omega_{n}\left|U_{n}\left(t\right)\right|\Omega_{n}\right\rangle =0
\end{eqnarray}
since $\left\langle \nu_{i};\mathbf{p}_{f},s_{f}\left|U_{i}\left(t\right)\right|\Omega_{i}\right\rangle =0$, as one can check with a direct calculation. Therefore we can write:
\begin{equation} \label{eq:cfhgfd}
T_{\alpha\rightarrow\beta} =\sum_{j=1}^{n}\text{U}_{\alpha j}\text{U}_{\beta j}^{*}\left[\left\langle \Omega_{1}\left|U_{1}\left(t\right)\right|\Omega_{1}\right\rangle ...\left\langle \nu_{j};\mathbf{p}_{f},s_{f}\left|U_{j}\left(t\right)\right|\nu_{j};\mathbf{p}_{i},s_{i}\right\rangle ....\left\langle \Omega_{n}\left|U_{n}\left(t\right)\right|\Omega_{n}\right\rangle \right],
\end{equation}
which reduces the entire calculation to a 1-particle computation.
In the next section we will focus our attention on the matrix element $\left\langle \nu_{j};\mathbf{p}_{f},s_{f}\left|U_{j}\left(t\right)\right|\nu_{j};\mathbf{p}_{i},s_{i}\right\rangle$ since, as we will show, the remaining terms contribute with an unimportant global phase factor.

\section{The matrix elements}
We now focus on the main part of this work. What we need to compute the 1-particle matrix element:
\begin{equation}
T\left(\mathbf{p}_{f},s_{f};\mathbf{p}_{i},s_{i};t\right)
\equiv
\left\langle \mathbf{p}_{f},s_{f}\left|U\left(t\right)\right|\mathbf{p}_{i},s_{i}\right\rangle = e^{-\frac{i}{\hbar}E_{f}t}\left\langle \mathbf{p}_{f},s_{f}\left|U_{I}\left(t\right)\right|\mathbf{p}_{i},s_{i}\right\rangle. 
\end{equation}
Since this part of the computation is the same for every mass eigenstate, we have dropped the label $j$. We expand the evolution operator by means of the Dyson series up to the second order:
\begin{equation}
U_{I}\left(t,0\right)\simeq1-\frac{i}{\hbar}\int_{0}^{t}dt_{1}:N_{I}\left(t_{1}\right):-\frac{1}{\hbar^{2}}\int_{0}^{t}dt_{1}\int_{0}^{t_{1}}dt_{2}:N_{I}\left(t_{1}\right)::N_{I}\left(t_{2}\right):,
\end{equation}
where $N_{I}\left(t\right)$ is the interaction picture representation of Eq.~(\ref{eq:dfgdf}) and $: ... :$ denotes the normal ordering\footnote{As well know in Quantum Field Theory, the reason why we used $:N_{I}\left(t\right):$ in place of $N_{I}\left(t\right)$ is that, with this prescription, we can remove all divergent contributions coming from tadpole diagrams. This type of divergences can be absorbed through a renormalization procedure, without giving any physically observable consequence~\cite{GreinerQED,GreinerFIELD}.}.
Accordingly, the transition probability is the sum of three terms:
\begin{equation} \label{eq:uioi}
T\left(\mathbf{p}_{f},s_{f};\mathbf{p}_{i},s_{i};t\right)=e^{-\frac{i}{\hbar}E_{f}t}\left[T^{\left(0\right)}\left(\mathbf{p}_{f},s_{f};\mathbf{p}_{i},s_{i};t\right)+T^{\left(1\right)}\left(\mathbf{p}_{f},s_{f};\mathbf{p}_{i},s_{i};t\right)+T^{\left(2\right)}\left(\mathbf{p}_{f},s_{f};\mathbf{p}_{i},s_{i};t\right)\right],
\end{equation}
corresponding to the first three terms of the Dyson series. We now give a pictorial representation of each term by means of Feynman diagram and we compute each of them. The first term corresponds to the free propagation:
\[
\begin{array}{cclcclccl}
\displaystyle  T^{\left(0\right)}\left(\mathbf{p}_{f},s_{f};\mathbf{p}_{i},s_{i};t\right) & = & \Diagram{\vertexlabel^{i}fsfAfs\vertexlabel^{f}}
\end{array}
\]   
where the solid line represent the particle. This term is trivial:
\begin{equation} \label{eq:cvhjcg}
T^{\left(0\right)}\left(\mathbf{p}_{f},s_{f};\mathbf{p}_{i},s_{i};t\right)
\equiv
\left\langle \mathbf{p}_{f},s_{f}|\mathbf{p}_{i},s_{i}\right\rangle =\delta_{s_{f}s_{i}}\delta_{\mathbf{p}_{f},\mathbf{p}_{i}}.
\end{equation}
The second term correspond to the diagram:
\[
\begin{array}{cclcclccl}
\displaystyle T^{\left(1\right)}\left(\mathbf{p}_{f},s_{f};\mathbf{p}_{i},s_{i};t\right) & = & \Diagram{\vertexlabel^ifA\vertexlabel_1 hu \\ f0 fdA\vertexlabel^{\;\;f}} 
\end{array}
\]  
where the dotted line represents the noise field. This term is:
\begin{equation}\label{eq:T1}
T^{\left(1\right)}\left(\mathbf{p}_{f},s_{f};\mathbf{p}_{i},s_{i};t\right)
\equiv
i\sqrt{\gamma_{m}}\int_{0}^{t}dt_{1}\int d\mathbf{x}_{1}w\left(x_{1}\right)\langle \mathbf{p}_{f},s_{f}|:\bar{\psi}_{I}\left(x_{1}\right)\psi_{I}\left(x_{1}\right):|\mathbf{p}_{i},s_{i}\rangle.
\end{equation}
In order to compute the matrix element in Eq.~(\ref{eq:T1}), we use the series expansion of the fields as given in Eq.~(\ref{eq:wave}). The non-null terms are those containing two $b$ and two $b^{\dagger}$ operators. After some calculations, one finds that:
\begin{eqnarray}\label{eq:T1me}
\lefteqn{\langle \Omega|b\left(p_{f},s_{f}\right):\bar{\psi}_{I}\left(x_{1}\right)\psi_{I}\left(x_{1}\right):b^{\dagger}\left(p_{i},s_{i}\right)|\Omega\rangle =} \nonumber  \\
& = & \sum_{s,s'=1}^{2}\sum_{\mathbf{p},\mathbf{p'}=-\infty}^{+\infty}\frac{1}{L^{3}}\frac{mc^{2}}{\sqrt{E_{p}E_{p'}}}e^{\frac{i}{\hbar}\left(p'^{\mu}-p^{\mu}\right)x_{1\mu}}\overline{u}\left(p',s'\right)u\left(p,s\right)\langle \Omega|b\left(p_{f},s_{f}\right)b^{\dagger}\left(p',s'\right)b\left(p,s\right)b^{\dagger}\left(p_{i},s_{i}\right)|\Omega\rangle \nonumber \\
& = & \frac{1}{L^{3}}\frac{mc^{2}}{\sqrt{E_{i}E_{f}}}e^{\frac{i}{\hbar}\left(p_{f}^{\mu}-p_{i}^{\mu}\right)x_{1\mu}}\overline{u}\left(p_{f},s_{f}\right)u\left(p_{i},s_{i}\right). 
\end{eqnarray}
Here we introduced the four momentum $p^{\mu}=\left(E_{p}/c,\mathbf{p}\right)$.
If we substitute Eq.~(\ref{eq:T1me}) in the definition of $T^{\left(1\right)}$, we get:
\begin{equation}\label{eq:T1again}
T^{\left(1\right)}\left(\mathbf{p}_{f},s_{f};\mathbf{p}_{i},s_{i};t\right)=i\sqrt{\gamma_{m}}\frac{mc^{2}}{\sqrt{E_{i}E_{f}}}\overline{u}\left(p_{f},s_{f}\right)u\left(p_{i},s_{i}\right)\frac{1}{L^{3}}\int_{0}^{t}dt_{1}\int d\mathbf{x}_{1}w\left(x_{1}\right)e^{\frac{i}{\hbar}\left(p_{f}^{\mu}-p_{i}^{\mu}\right)x_{1\mu}}.
\end{equation}

The last term in Eq.~(\ref{eq:uioi}) is the more complicated to compute, since it involves the product of four fields. It gives the following contribution:
\begin{eqnarray}
\lefteqn{T^{\left(2\right)}\left(\mathbf{p}_{f},s_{f};\mathbf{p}_{i},s_{i};t\right) 
\equiv
-\frac{1}{\hbar^{2}}\langle \mathbf{p}_{f},s_{f}|\left[\int_{0}^{t}dt_{1}\int_{0}^{t_{1}}dt_{2}:N_{I}\left(t_{1}\right)::N_{I}\left(t_{2}\right):\right]|\mathbf{p}_{i},s_{i}\rangle =}  \\
& = & -\frac{\gamma_{m}}{2}\int_{0}^{t}\!dt_{1}dt_{2}\int\! d\mathbf{x}_{1} d\mathbf{x}_{2}\,w\left(x_{1}\right)w\left(x_{2}\right) \langle \mathbf{p}_{f},s_{f}|T\left[:\bar{\psi}_{I}\left(x_{1}\right)\psi_{I}\left(x_{1}\right)::\bar{\psi}_{I}\left(x_{2}\right)\psi_{I}\left(x_{2}\right):\right]|\mathbf{p}_{i},s_{i}\rangle, \nonumber
\end{eqnarray}
here ``$T$'' is the time-ordering product. Using Wick's theorem, and discarding all tadpole terms, which involve a contraction between two fields at the same spacetime point, we have:
\begin{eqnarray} \label{eq:dsfdfbv}
T\left[\bar{\psi}_{1a}\psi_{1a}\bar{\psi}_{2b}\psi_{2b}\right]
& = & :\bar{\psi}_{1a}\psi_{1a}\bar{\psi}_{2b}\psi_{2b}:-S_{ab}\left(x_{1}-x_{2}\right)S_{ba}\left(x_{2}-x_{1}\right) \nonumber \\
& & +iS_{ab}\left(x_{1}-x_{2}\right):\bar{\psi}_{1a}\psi_{2b}:-iS_{ba}\left(x_{2}-x_{1}\right):\psi_{1a}\bar{\psi}_{2b}:,
\end{eqnarray}
where $a$ and $b$ label the spinor components and the Dirac propagator is: $ iS_{ab}\left(x_{1}-x_{2}\right)\equiv\langle \Omega|T\left[\psi_{1a}\bar{\psi}_{2b}\right]|\Omega\rangle$. Here, Einstein's summation convention is used for the spinor indices. We momentarily drop the pedex $I$ related to the interaction picture and we write the dependence on $x_{1}$ and $x_{2}$ simply as $_{1}$ and $ _{2}$. The diagramatic representation of the different terms in Eq.~(\ref{eq:dsfdfbv}) is:
\[
\begin{array}{cclcclccl}
\Diagram{\vertexlabel^ifA\vertexlabel_1 hu \\ f0 fdA\vertexlabel^{\;\;f}}\;\;\;\;\Diagram{\vertexlabel^ifA\vertexlabel_2 hu \\ f0 fdA\vertexlabel^{\;\;f}} & = & \displaystyle :\bar{\psi}_{1a}\psi_{1a}\bar{\psi}_{2b}\psi_{2b}:\;, \qquad &
\feyn{h\vertexlabel_1 f0 {flA}{}
{fluV}{} f0 \vertexlabel_2h}  & = & \displaystyle -S_{ab}\left(x_{1}-x_{2}\right)S_{ba}\left(x_{2}-x_{1}\right)\;,
\end{array}
\]
\[
\begin{array}{cclcclccl}
\Diagram{\vertexlabel_i hd\vertexlabel_{2}  fA \vertexlabel_{1}
hu\\
fuA f0 fdA \vertexlabel^{\;\;f}}  & = & \displaystyle iS_{ab}\left(x_{1}-x_{2}\right):\bar{\psi}_{1a}\psi_{2b}:\;,\qquad\qquad\qquad
\Diagram{\vertexlabel_i hd\vertexlabel_{1}  fA \vertexlabel_{2}
hu\\
fuA f0 fdA \vertexlabel^{\;\;f}}  & = & \displaystyle -iS_{ba}\left(x_{2}-x_{1}\right):\psi_{1a}\bar{\psi}_{2b}:\;.
\end{array}
\]
We can easily see that the first term is zero, since we are studying the case with only one particle in the initial and final states. Regarding the second diagram, an important issue here arises. This diagam represents a vacuum fluctuation term, which is divergent. As well known~\cite{GreinerFIELD}, all vacuum fluctuations diagrams of any order sum up to a phase factor $\left\langle \Omega\left|U_{I}\left(t,0\right)\right|\Omega\right\rangle$, and all divergences cancel each other. Therefore we can write:
\begin{equation}
\left\langle \mathbf{p}_{f},s_{f}\left|U_{I}\left(t,0\right)\right|\mathbf{p}_{i},s_{i}\right\rangle =\left\langle \Omega\left|U_{I}\left(t,0\right)\right|\Omega\right\rangle \cdot \left\langle \mathbf{p}_{f},s_{f}\left|U_{I}\left(t,0\right)\right|\mathbf{p}_{i},s_{i}\right\rangle _{\textrm{ext}},
\end{equation}
where $\left\langle \mathbf{p}_{f},s_{f}\left|U_{I}\left(t,0\right)\right|\mathbf{p}_{i},s_{i}\right\rangle_{\textrm{ext}}$ denotes the contribution from diagrams with external fermionic legs. This vacuum fluctuation term is important because, together with those of Eq.~(\ref{eq:cfhgfd}), it gives a global phase $\prod_{k=1}^{n}\left\langle \Omega_{k}\left|U_{kI}\left(t,0\right)\right|\Omega_{k}\right\rangle$ independent of $j$, which factorizes out of the sum. Therefore, such terms are physically unimportant, the only relevant part being $\left\langle \mathbf{p}_{f},s_{f}\left|U_{I}\left(t,0\right)\right|\mathbf{p}_{i},s_{i}\right\rangle_{\textrm{ext}}$. From now on, we will work only with diagrams with external fermionic legs, and we drop the pedex ``ext''. 

Coming back to Eq.~(\ref{eq:dsfdfbv}), we can now focus our attention on the third and the fourth term, that correspond to the last two diagrams. Since $:\psi_{2b}\bar{\psi}_{1a}:\; = - :\bar{\psi}_{1a}\psi_{2b}:$ for fermions, these two terms give the same contribution. The Dirac propagator reads:
\begin{eqnarray}
iS_{ab}\left(x_{1}-x_{2}\right)
& =& 
\sum_{s=1}^{2}\sum_{\mathbf{p}=-\infty}^{+\infty}\frac{1}{L^{3}}\frac{mc^{2}}{E_{p}}\left\{ \theta\left(t_{1}-t_{2}\right)e^{-\frac{i}{\hbar}p^{\mu}\left(x_{1\mu}-x_{2\mu}\right)}u_{a}\left(p,s\right)\overline{u}_{b}\left(p,s\right) \right. \nonumber \\
& & \left.\qquad\qquad\qquad\quad\;\; -\theta\left(t_{2}-t_{1}\right)e^{-\frac{i}{\hbar}p^{\mu}\left(x_{2\mu}-x_{1\mu}\right)}\overline{v}_{b}\left(p,s\right)v_{a}\left(p,s\right)\right\}, 
\end{eqnarray}
while the matrix element gives:
\begin{equation}
\langle \mathbf{p}_{f},s_{f}|:\bar{\psi}_{1a}\psi_{2b}:|\mathbf{p}_{i},s_{i}\rangle 
=
\frac{1}{L^{3}}\frac{mc^{2}}{\sqrt{E_{f}E_{i}}}e^{\frac{i}{\hbar}p_{f}^{\mu}x_{1\mu}}e^{-\frac{i}{\hbar}p_{i}^{\mu}x_{2\mu}}\overline{u}_{a}\left(p_{f},s_{f}\right)u_{b}\left(p_{i},s_{i}\right).
\end{equation}
In the following we will need only the case $\mathbf{p}_{f}=\mathbf{p}_{i}$ and $s_{f}=s_{i}$. In this case, we have:
\begin{eqnarray}
T^{\left(2\right)}\left(\mathbf{p}_{i},s_{i};\mathbf{p}_{i},s_{i};t\right)
& = & 
-\gamma_{m}\frac{1}{L^{3}}\frac{m c^{2}}{E_{i}}\sum_{s=1}^{2}\sum_{\mathbf{p}=-\infty}^{+\infty}\frac{1}{L^{3}}\frac{m c^{2}}{E_{p}} \nonumber \\
& & 
\cdot\left\{ \int_{0}^{t}dt_{1}\int_{0}^{t_{1}}dt_{2}\int d\mathbf{x}_{1}\int d\mathbf{x}_{2}\, w(x_{1})w(x_{2})e^{\frac{i}{\hbar}(p_{i}^{\mu}-p^{\mu})\left(x_{1\mu}-x_{2\mu}\right)}\right. \nonumber \\
& & \cdot \; \overline{u}_{a}(p_{i},s_{i})u_{a}(p^,s)\overline{u}_{b}(p^,s) u_{b}(p_{i},s_{i}) \nonumber \\
& &
-\int_{0}^{t}dt_{2}\int_{0}^{t_{2}}dt_{1}\int d\mathbf{x}_{1}\int d\mathbf{x}_{2}w\left(x_{1}\right)w\left(x_{2}\right)e^{\frac{i}{\hbar}(p_{i}^{\mu}-p^{\mu})\left(x_{1\mu}-x_{2\mu}\right)} \nonumber \\
& & \cdot \left.
\overline{u}_{a}(p_{i},s_{i})v_{a}(p,s)\overline{v}_{b}(p,s)u_{b}(p_{i},s_{i})\phantom{\frac{1}{2}}\!\!\!\!\right\}.
\end{eqnarray}
Using the standard relations~\cite{Greiner}:
\begin{equation}
\sum_{s=1}^{2}u_{a}(p,s)\overline{u}_{b}(p,s)=\left(\frac{p^{\mu}\gamma_{\mu}+mc}{2mc}\right)_{ab},\;\;\;\;\;\;\sum_{s=1}^{2}v_{a}(p,s)\overline{v}_{b}(p,s)=\left(\frac{p^{\mu}\gamma_{\mu}-mc}{2mc}\right)_{ab},
\end{equation}
we can see that the terms containing $p^{\mu}\gamma_{\mu}$ cancel each other, while those containing the mass give a $\delta_{ab}$. Thus, if we also use: $\overline{u}(p_{i},s_{f})u(p_{i},s_{i})=\delta_{s_{f},s_{i}}$, we obtain:
\begin{eqnarray} \label{eq:fdhgfg}
T^{\left(2\right)}\left(\mathbf{p}_{i},s_{i};\mathbf{p}_{i},s_{i};t\right)
& = &
-\gamma_{m}\frac{1}{L^{3}}\frac{mc^{2}}{E_{i}}\sum_{\mathbf{p}=-\infty}^{+\infty}\frac{1}{L^{3}}\frac{mc^{2}}{E_{p}}\int_{0}^{t}dt_{1}\int_{0}^{t_{1}}dt_{2}\int d\mathbf{x}_{1}\int d\mathbf{x}_{2}w\left(x_{1}\right)w\left(x_{2}\right) \nonumber \\
& & \times e^{\frac{i}{\hbar}(p_{i}^{\mu}-p^{\mu})\left(x_{1\mu}-x_{2\mu}\right)}.
\end{eqnarray}
Now we have all the elements we need, in order to compute the transition probability. We will do this in the next section.

\section{The transition probability}

The physical quantity we are interested in, is the transition probability, which corresponds to $\left|T_{\alpha\rightarrow\beta}\right|^{2}$, averaged over the noise, and integrated over the final momentum and polarization states: 
\begin{equation} \label{eq:sfdso}
P_{\alpha\rightarrow\beta}
\equiv
\sum_{s_{f}}\sum_{\mathbf{p}_{f}=-\infty}^{+\infty}\mathbb{E}\left|T_{\alpha\rightarrow\beta}\right|^{2}
=
\sum_{k=1}^{n}\sum_{j=1}^{n}\text{U}_{\alpha k}^{*}\text{U}_{\beta k}\text{U}_{\alpha j}\text{U}_{\beta j}^{*}P_{kj}\left(\mathbf{p}_{i},s_{i};t\right),
\end{equation}
where:
\begin{equation}
P_{kj}\left(\mathbf{p}_{i},s_{i};t\right)\equiv\sum_{s_{f}}\sum_{\mathbf{p}_{f}=-\infty}^{+\infty}e^{\frac{i}{\hbar}(E_{f}^{\left(k\right)}-E_{f}^{\left(j\right)})t}\,\mathbb{E}\left[T_{k}^{*}\left(\mathbf{p}_{f},s_{f};\mathbf{p}_{i},s_{i};t\right) T_{j}\left(\mathbf{p}_{f},s_{f};\mathbf{p}_{i},s_{i};t\right)\right],
\end{equation}
and $T_{j}\left(\mathbf{p}_{f},s_{f};\mathbf{p}_{i},s_{i};t\right)$ is given by Eq.~(\ref{eq:uioi}), where now we have explicitly indicated the label $j$ associated to the mass eigenstate $m_{j}$ and $E_{f}^{\left(j\right)}=\sqrt{\mathbf{p}_f^{2}c^{2}+m_j^{2}c^{4}}$. When averaging, one has to remember that only terms containing an even number of noises survive (in the Feynman representation, all products of diagrams containing an even number of dotted legs). Using this fact, and exploiting the Kronecher deltas of $T^{\left(0\right)}\left(\mathbf{p}_{f},s_{f};\mathbf{p}_{i},s_{i};t\right)$ (see Eq.~(\ref{eq:cvhjcg})), we can write:  
\begin{equation} \label{eq:sdgrst}
P_{kj}\left(\mathbf{p}_{i},s_{i};t\right)
=
e^{\frac{i}{\hbar}(E_{i}^{\left(k\right)}-E_{i}^{\left(j\right)})t}\left[1+ I_{jk}^{\left(1\right)}\left(\mathbf{p}_{i},s_{i};t\right) + I_{j}^{\left(2\right)}\left(\mathbf{p}_{i},s_{i};t\right)+I_{k}^{\left(2\right)*}\left(\mathbf{p}_{i},s_{i};t\right)\right],
\end{equation}
where we have defined:
\begin{eqnarray} \label{eq:rtrtsfds}
I_{jk}^{\left(1\right)}\left(\mathbf{p}_{i},s_{i};t\right)
& \equiv &
\sum_{s_{f}}\sum_{\mathbf{p}_{f}=-\infty}^{+\infty}e^{\frac{i}{\hbar}(E_{f}^{\left(k\right)}-E_{i}^{\left(k\right)}-E_{f}^{\left(j\right)}+E_{i}^{\left(j\right)})t}\mathbb{E}\left[T_{k}^{\left(1\right)*}\left(\mathbf{p}_{f},s_{f};\mathbf{p}_{i},s_{i};t\right)T_{j}^{\left(1\right)}\left(\mathbf{p}_{f},s_{f};\mathbf{p}_{i},s_{i};t\right)\right], \nonumber \\
I_{j}^{\left(2\right)}\left(\mathbf{p}_{i},s_{i};t\right)
& \equiv & 
\mathbb{E}\left[T_{j}^{\left(2\right)}\left(\mathbf{p}_{i},s_{i};\mathbf{p}_{i},s_{i};t\right)\right].
\end{eqnarray}
We focus our attention on $I_{jk}^{\left(1\right)}\left(\mathbf{p}_{i},s_{i};t\right)$. Using Eq.~(\ref{eq:T1again}), keeping in mind the spinor relation $\left(\overline{u}_{f}u_{i}\right)^{*}=\overline{u}_{i}u_{f}$, and performing the average over the noise, which brings in a Dirac delta in time which cancels one of the two time-integrals, one obtains:
\begin{eqnarray} \label{eq:sdfgdr}
\lefteqn{\mathbb{E}\left[T_{k}^{\left(1\right)*}\left(\mathbf{p}_{f},s_{f};\mathbf{p}_{i},s_{i};t\right)T_{j}^{\left(1\right)}\left(\mathbf{p}_{f},s_{f};\mathbf{p}_{i},s_{i};t\right)\right]=} \qquad\qquad\nonumber \\
& = & 
\sqrt{\gamma_{m_{j}}\gamma_{m_{k}}}\frac{m_{j}m_{k}c^{4}}{\sqrt{E_{i}^{\left(j\right)}E_{f}^{\left(j\right)}E_{i}^{\left(k\right)}E_{f}^{\left(k\right)}}}\overline{u}(p_{f}^{(j)},s_{f})u(p_{i}^{(j)},s_{i})\overline{u}(p_{i}^{(k)},s_{i})u(p_{f}^{(k)},s_{f}) \nonumber \\
& & 
\cdot  \int_{0}^{t}dt_{1}e^{\frac{i}{\hbar}(E_{f}^{\left(j\right)}-E_{i}^{\left(j\right)}-E_{f}^{\left(k\right)}+E_{i}^{\left(k\right)})t_{1}}S\left(\mathbf{p}_{i},\mathbf{p}_{f}\right)
\end{eqnarray}
where $p_{f}^{(j)} \equiv (E_{f}^{(j)}/c, {\bf p}_{f})$ and similarly for $p_{i}^{(j)}$. Moreover:
\begin{equation} \label{eq:zxgf}
S\left(\mathbf{p}_{i},\mathbf{p}_{f}\right)
\equiv
\frac{1}{L^{6}}\int_{-\frac{L}{2}}^{+\frac{L}{2}}d\mathbf{x}_{1}\int_{-\frac{L}{2}}^{+\frac{L}{2}}d\mathbf{x}_{2}\frac{e^{-\left(\mathbf{x}_{1}-\mathbf{x}_{2}\right)^{2}/4r_{C}^{2}}}{\left(\sqrt{4\pi}r_{C}\right)^{3}}e^{-\frac{i}{\hbar}\left[\left(\mathbf{p}_{f}-\mathbf{p}_{i}\right)\cdot\left(\mathbf{x}_{1}-\mathbf{x}_{2}\right)\right]}
\end{equation}
(now we have explicitly indicated the integration volume). In order to compute $S\left(\mathbf{p}_{i},\mathbf{p}_{f}\right)$, we change integration variables as follows:
\begin{equation}
\mathbf{y}=\left(\mathbf{x}_{1}+\mathbf{x}_{2}\right)\qquad\textrm{and}\qquad\mathbf{x}=\left(\mathbf{x}_{1}-\mathbf{x}_{2}\right),
\end{equation}
and use the relation:
\begin{equation}
\int_{-\frac{L}{2}}^{+\frac{L}{2}}dx_{1}\int_{-\frac{L}{2}}^{+\frac{L}{2}}dx_{2}f\left(x_{1},x_{2}\right)=\frac{1}{2}\int_{0}^{+L}dx\int_{-\left(L-x\right)}^{+\left(L-x\right)}dy\left[f\left(x,y\right)+f\left(-x,y\right)\right].
\end{equation}
Accordingly, we have:
\begin{equation}
S\left(\mathbf{p}_{i},\mathbf{p}_{f}\right)=\frac{1}{L^{3}}\int_{0}^{L}d\mathbf{x}\frac{e^{-\mathbf{x}^{2}/4r_{C}^{2}}}{\left(\sqrt{4\pi}r_{C}\right)^{3}}2\cos\left[\frac{1}{\hbar}\left(\mathbf{p}_{f}-\mathbf{p}_{i}\right)\cdot\mathbf{x}\right]\frac{1}{2^{3}}\prod_{i=1}^{3}2\left(1-\frac{x_{i}}{L}\right)
\end{equation}
Let us now take the limit $L \rightarrow \infty$, which amounts to making the replacement:
\begin{equation}
\sum_{\mathbf{p}_{f}=-\infty}^{+\infty}\longrightarrow\int d\mathbf{p}_{f}\qquad\textrm{and}\qquad\frac{1}{L^{3}}\longrightarrow\frac{1}{\left(2\pi\hbar\right)^{3}}.
\end{equation}
In this limit, the term $x_i/L$ gives a vanishingly small contribution. Therefore we can write:
\begin{equation}
S\left(\mathbf{p}_{i},\mathbf{p}_{f}\right)
=
\frac{1}{\left(2\pi\hbar\right)^{3}} \int_{-\infty}^{+\infty}d\mathbf{x}\frac{e^{-\mathbf{x}^{2}/4r_{C}^{2}}}{\left(\sqrt{4\pi}r_{C}\right)^{3}}e^{\frac{i}{\hbar}(\mathbf{p}_{f}-\mathbf{p}_{i})\cdot\mathbf{x}} 
= 
\frac{1}{\left(2\pi\hbar\right)^{3}} e^{-\frac{(\mathbf{p}_{f}-\mathbf{p}_{i})^{2}r_{C}^{2}}{\hbar^{2}}}.
\end{equation}
The time integral in Eq.~(\ref{eq:sdfgdr}) is trivial, and one arrives easily at the formula:
\begin{eqnarray} \label{eq:dasdasd}
I_{jk}^{\left(1\right)}\left(\mathbf{p}_{i},s_{i};t\right)
& = & 
\sum_{s_{f}}\int d\mathbf{p}_{f}\sqrt{\gamma_{m_{j}}\gamma_{m_{k}}}\frac{m_{j}m_{k}c^{4}}{\sqrt{E_{i}^{\left(j\right)}E_{f}^{\left(j\right)}E_{i}^{\left(k\right)}E_{f}^{\left(k\right)}}}\overline{u}(p_{f}^{\left(j\right)},s_{f})u(p_{i}^{\left(j\right)},s_{i})\overline{u}(p_{i}^{\left(k\right)},s_{i})u(p_{f}^{\left(k\right)},s_{f}) \nonumber \\
& & \cdot
\frac{1-e^{\frac{i}{\hbar}(E_{f}^{\left(k\right)}-E_{i}^{\left(k\right)}-E_{f}^{\left(j\right)}+E_{i}^{\left(j\right)})t}}{\frac{i}{\hbar}(E_{f}^{\left(j\right)}-E_{i}^{\left(j\right)}-E_{f}^{\left(k\right)}+E_{i}^{\left(k\right)})}\frac{1}{\left(2\pi\hbar\right)^{3}}e^{-\frac{(\mathbf{p}_{f}-\mathbf{p}_{i})^{2}r_{C}^{2}}{\hbar^{2}}}.
\end{eqnarray}
As it is shown in the Appendix C, the integrating function (except for the Gaussian term) changes slowly within the region where the Gaussian term is appreciably different from zero. Therefore we can approximate it with the value it takes in the center of the Gaussian (where $\mathbf{p}_{f}=\mathbf{p}_{i}$) and bring it out of the integral. Taking into account that $\overline{u}\left(p,s_{i}\right)u\left(p,s_{f}\right)=\delta_{s_{i}s_{f}}$ and performing the integration of the Gaussian part, Eq.~(\ref{eq:dasdasd}) takes the very simple expression:
\begin{equation} \label{eq:syksdgar}
I_{jk}^{\left(1\right)}\left(\mathbf{p}_{i},s_{i};t\right)
=
\sqrt{\gamma_{m_{j}}\gamma_{m_{k}}}\frac{m_{j}m_{k}c^{4}}{E_{i}^{\left(j\right)}E_{i}^{\left(k\right)}}\frac{t}{\left(2\pi\right)^{3}}\frac{\pi^{3/2}}{r_{C}^{3}}.
\end{equation}

We now turn our attention to the term $I_{j}^{\left(2\right)}\left(\mathbf{p}_{i},s_{i};t\right)$ in Eq.~(\ref{eq:rtrtsfds}). Substituting Eq.~(\ref{eq:fdhgfg}) in Eq.~(\ref{eq:rtrtsfds}), we have: 
\begin{equation}
I_{j}^{\left(2\right)}\left(\mathbf{p}_{i},s_{i};t\right)
=
-\gamma_{m_{j}}\frac{m_{j}c^{2}}{E_{i}^{\left(j\right)}}\sum_{\mathbf{p}=-\infty}^{+\infty}\frac{m_{j}c^{2}}{E_{p}^{\left(j\right)}}\frac{t}{2}\frac{1}{L^{6}}\int d\mathbf{x}_{1}\int d\mathbf{x}_{2}\frac{e^{-\left(\mathbf{x}_{1}-\mathbf{x}_{2}\right)^{2}/4r_{C}^{2}}}{\left(\sqrt{4\pi}r_{C}\right)^{3}}e^{-\frac{i}{\hbar}\left(\mathbf{p}_{i}-\mathbf{p}\right)\left(\mathbf{x}_{1}-\mathbf{x}_{2}\right)}.
\end{equation}
The spatial integrals are equal to $S\left(\mathbf{p},\mathbf{p}_{i}\right)$ defined in Eq.~(\ref{eq:zxgf}). Performing the same type of calculation as before, and taking the limit $L \rightarrow \infty$, one arrives at the result:
\begin{equation} \label{eq:vcvxvxcy}
I_{j}^{\left(2\right)}\left(\mathbf{p}_{i},s_{i};t\right)
=
-\gamma_{m_{j}}\frac{m_{j}c^{2}}{E_{i}^{\left(j\right)}}\int d\mathbf{p}\frac{m_{j}c^{2}}{E_{p}^{\left(j\right)}}\frac{1}{2}t\frac{1}{\left(2\pi\hbar\right)^{3}}e^{-\frac{\left(\mathbf{p}-\mathbf{p}_{i}\right)^{2}r_{C}^{2}}{\hbar^{2}}}.
\end{equation}
Once again, one can show that the integrating function (Gaussian term apart) varies slowly within the region where the Gaussian function is appreciably different from zero. Therefore one can bring this function out of the integral, fixing its value at the center of the Gaussian ($\mathbf{p}=\mathbf{p}_{i}$), and perform the Gaussian integration. The final expression is:
\begin{equation} \label{eq:izgdj}
I_{j}^{\left(2\right)}\left(\mathbf{p}_{i},s_{i};t\right)
=
-\frac{\gamma_{m_{j}}}{2}\frac{m_{j}^{2}c^{4}}{E_{i}^{2\left(j\right)}}\frac{t}{\left(2\pi\right)^{3}}\frac{\pi^{3/2}}{r_{C}^{3}}.
\end{equation}

Having computed explicitly all the terms of Eq.~(\ref{eq:sdgrst}), we can turn our attention to Eq.~(\ref{eq:sfdso}). It is convenient to split the sum of Eq.~(\ref{eq:sfdso}) in one part with $k=j$ and the another part with $k\neq j$:
\begin{eqnarray}\label{eq:ennesimaP}
P_{\alpha\rightarrow\beta}
& = & 
\sum_{k=1}^{n}\text{U}_{\alpha k}^{*}\text{U}_{\beta k}\text{U}_{\alpha k}\text{U}_{\beta k}^{*}P_{kk}\left(\mathbf{p}_{i},s_{i};t\right) \nonumber \\
& + &
\sum_{{k=2\atop j<k}}^{n}\left[\text{U}_{\alpha k}^{*}\text{U}_{\beta k}\text{U}_{\alpha j}\text{U}_{\beta j}^{*}P_{kj}\left(\mathbf{p}_{i},s_{i};t\right)+\text{U}_{\alpha j}^{*}\text{U}_{\beta j}\text{U}_{\alpha k}\text{U}_{\beta k}^{*}P_{jk}\left(\mathbf{p}_{i},s_{i};t\right)\right]\,.
\end{eqnarray}
Using the symmetry relation $I_{jk}^{\left(1\right)}=I_{kj}^{\left(1\right)*}$, which implies that $P_{kj}\left(\mathbf{p}_{i},s_{i};t\right)=P_{jk}^{*}\left(\mathbf{p}_{i},s_{i};t\right)$, one can rewrite Eq.~\eqref{eq:ennesimaP} as follows :
\begin{equation} \label{eq:dgglkjgh}
P_{\alpha\rightarrow\beta}=\sum_{k=1}^{n}\text{U}_{\alpha k}^{*}\text{U}_{\beta k}\text{U}_{\alpha k}\text{U}_{\beta k}^{*}+\sum_{{k=2\atop j<k}}^{n}2\textrm{Re}\left[\text{U}_{\alpha k}^{*}\text{U}_{\beta k}\text{U}_{\alpha j}\text{U}_{\beta j}^{*}P_{kj}\left(\mathbf{p}_{i},s_{i};t\right)\right]\,,
\end{equation}
where we have exploited the identity:
$
P_{kk}\left(\mathbf{p}_{i},s_{i};t\right)=1.
$
In the physically interesting case where the mixing elements $U_{\alpha k}$ are real, Eq.~(\ref{eq:dgglkjgh}) takes a very simple expression:
\begin{equation} \label{eq:dfggkds}
P_{\alpha\rightarrow\beta}
=
\sum_{k=1}^{n}\text{U}_{\alpha k}\text{U}_{\beta k}\text{U}_{\alpha k}\text{U}_{\beta k}+\sum_{{k \neq j}}^{n}\text{U}_{\alpha k}\text{U}_{\beta k}\text{U}_{\alpha j}\text{U}_{\beta j} \; \left[1- \xi_{jk} t\right] \; \cos\left[\frac{1}{\hbar}(E_{i}^{\left(k\right)}-E_{i}^{\left(j\right)})t\right],
\end{equation}
with:
\begin{equation}\label{eq:xi1}
\xi_{jk} \; \equiv \; \frac{1}{16\pi^{3/2}r_{C}^{3}}\left(\sqrt{\gamma_{m_{j}}}\frac{m_{j}c^{2}}{E_{i}^{\left(j\right)}}-\sqrt{\gamma_{m_{k}}}\frac{m_{k}c^{2}}{E_{i}^{\left(k\right)}}\right)^{2}= \frac{\gamma}{16\pi^{3/2}r_{C}^{3}m_{0}^{2}c^{4}}\left(\frac{m_{j}^{2}c^{4}}{E_{i}^{\left(j\right)}}-\frac{m_{k}^{2}c^{4}}{E_{i}^{\left(k\right)}}\right)^{2}\,.
\end{equation}
As a check, one can easily see that the probability is conserved, i.e.:
\begin{equation}
\sum_{\beta}P_{\alpha\rightarrow\beta}=1.
\end{equation}
This is the result we wanted to arrive at. It shows that, also in collapse models, the number of particles is conserved, but the oscillations are damped according to Eq.~(\ref{eq:dfggkds}), by a factor equal to $\left[1- \xi_{jk} t\right] $. This is in perfect agreement\footnote{In general, spontaneous collapses and decoherence are described by similar master equations.} with well established results concerning the effect of decoherence on oscillatory systems like those here considered~\cite{Be,Adlerdeco}. Our calculation gives an analytical expression for the damping rate $\xi_{jk}$, as predicted by the mass-proportional CSL model. Note that the calculation has been carried out to second perturbative order, which means $\xi_{jk} t \ll 1$. Therefore, the fact that the probability in Eq.~(\ref{eq:dfggkds}) becomes negative for $\xi_{jk} t > 1$ is of no concern, because this range of times goes beyond the limits of validity of the present result. Actually, one can try to stretch the above result beyond the second perturbative order, and guess the following expression for the transition probability:
\begin{equation} \label{eq:dfggkdse1}
P_{\alpha\rightarrow\beta}
=
\sum_{k=1}^{n}\text{U}_{\alpha k}\text{U}_{\beta k}\text{U}_{\alpha k}\text{U}_{\beta k}+\sum_{{k \neq j}}^{n}\text{U}_{\alpha k}\text{U}_{\beta k}\text{U}_{\alpha j}\text{U}_{\beta j} \; e^{-\xi_{jk}t} \; \cos\left[\frac{1}{\hbar}(E_{i}^{\left(k\right)}-E_{i}^{\left(j\right)})t\right].
\end{equation}

The above results can be easily generalized to oscillatory systems, which decay in time. On the phenomenological level, one takes the decay into account by adding an imaginary term to the Hamiltonian:
\begin{equation}
H\longrightarrow H-\frac{i}{2}\Gamma.
\end{equation}
The calculation remains unaltered, and one arrives at the final result: 
\begin{eqnarray} \label{eq:dfggkdsss}
P_{\alpha\rightarrow\beta}
& = &
\sum_{k=1}^{n}\text{U}_{\alpha k}\text{U}_{\beta k}\text{U}_{\alpha k}\text{U}_{\beta k} \; e^{-\frac{\Gamma^{\left(k\right)}}{\hbar}t} \nonumber \\
& + & \sum_{{k=2\atop j<k}}^{n}\text{U}_{\alpha k}\text{U}_{\beta k}\text{U}_{\alpha j}\text{U}_{\beta j} \; \left[1- \xi_{jk} t\right] \; e^{-\frac{\Gamma^{\left(k\right)}+\Gamma^{\left(j\right)}}{2\hbar}t} \; 2\cos\left[\frac{1}{\hbar}(E_{i}^{\left(k\right)}-E_{i}^{\left(j\right)})t\right],
\end{eqnarray}
which generalizes Eq.~(\ref{eq:dfggkds}) to decaying particles.  This concludes our analysis.

\section{Acknowledgements}

The authors wish to thank S.L. Adler, M. Bahrami, A. Di Domenico and B. Hiesmayr for many useful and enjoyable conversations on this topic. They also acknowledge partial financial support from NANOQUESTFIT, INFN, the COST Action MP1006 ``Fundamental problems in Quantum Physics" and the John Templeton Foundation project ``Experimental and theoretical exploration of fundamental limits of quantum mechanics''. 

\section*{APPENDIX A: Dimensional estimate of the decay rate}

Here we wish to discuss about the possibility of estimate the predictions of the CSL model in neutrino oscillations by using dimensional analysis. Collapse models, as discussed in section IV, are described
by the same type of master equations as open quantum systems (which experience
decoherence due to interactions with an external environment). In
such a case it is well known that the effect of decoherence is to suppress
exponentially flavour oscillations. So it does not come as a surprise that for collapse models the effect is the same. Then one could try to
guess the decay rate with dimensional analysis, using the relevant constants and parameters of
the model. First of all is reasonable to suppose that the effect is
proportional to the strength of the noise $\gamma$. Moreover, since
we are using the mass proportional CSL model, for which $\gamma$ is replaced by $\gamma_{m_{j}}\equiv\gamma\left(\frac{m_{j}}{m_{0}}\right)^{2}$, one expects also a factor $m_{0}^{2}$ in the denominator. Because
$\left[\gamma\right]=\textrm{cm}^{3}\textrm{s}^{-1}$, and the decay
rate must have dimension $\textrm{s}^{-1}$, we need to introduce
terms with dimension $\textrm{cm}^{-3}$. Since the parameter $r_{C}$ has the dimension of a length, then it is natural to introduce an $r_{C}^{3}$ in the denominator. Finally, we need to introduce
terms with the dimension of a squared mass. The simplest choices are: 
\begin{equation}
\xi_{jk}^{(1)}\sim\frac{\gamma}{r_{C}^{3}m_{0}^{2}}\left(m_{j}-m_{k}\right)^{2}\;\;\;\;\;\textrm{or}\;\;\;\;\;\xi_{jk}^{(2)}\sim\frac{\gamma}{r_{C}^{3}m_{0}^{2}}\left(m_{j}^{2}-m_{k}^{2}\right).
\end{equation}
Both these formulas are different compared to the correct one given
by Eq.~(\ref{eq:xi}). If we substitute the values of the constants and the parameters and we consider, for example, the case of the cosmological neutrinos, we get:
\begin{eqnarray}
\xi_{jk}^{(1)}t_{cosm}&\sim&\frac{\gamma}{r_{C}^{3}m_{0}^{2}}\left(m_{j}-m_{k}\right)^{2}t_{cosm}\sim10^{-17}\\
&& \nonumber\\
\xi_{jk}^{(2)}t_{cosm}&\sim&\frac{\gamma}{r_{C}^{3}m_{0}^{2}}\left(m_{j}^{2}-m_{k}^{2}\right)t_{cosm}\sim10^{-11}
\end{eqnarray}
The formula derived with dimensional analysis shows that the CSL effect on neutrino oscillations is very small, practically undetectable. However, it differs by many orders of magnitude from the exact (perturbative) result. We performed the lengthy calculation present here in order to arrive at a fully trustable result. As we have seen here above, dimensional analysis does not allow to reach a firm conclusion.  

\section*{APPENDIX B: Approximation in the calculation of $I_{j}^{\left(2\right)}$}

In this appendix we justify the approximation we used in order to derive Eq.~(\ref{eq:izgdj}) from Eq.~(\ref{eq:vcvxvxcy}). This amounts to proving that:
\begin{equation}
\frac{1}{\hbar^{3}}\int d\mathbf{p}\frac{1}{E_{p}^{\left(j\right)}}e^{-\frac{\left(\mathbf{p}-\mathbf{p}_{i}\right)^{2}r_{C}^{2}}{\hbar^{2}}}\simeq\frac{1}{E_{i}^{\left(j\right)}}\frac{\pi^{3/2}}{r_{C}^{3}}
\end{equation}
To see this, we can rewrite the integral in polar coordinates and integrate over the angular variables:
\begin{equation}
\frac{1}{\hbar^{3}}\int d\mathbf{p}\frac{1}{E_{p}^{\left(j\right)}}e^{-\frac{\left(\mathbf{p}-\mathbf{p}_{i}\right)^{2}r_{C}^{2}}{\hbar^{2}}}=\frac{2\pi}{\hbar^{3}}\frac{\hbar^{2}}{2p_{i}r_{C}^{2}}\int_{-\infty}^{+\infty}dp\frac{p}{\sqrt{p^{2}c^{2}+m_{j}^{2}c^{4}}}e^{-\frac{\left(p-p_{i}\right)^{2}r_{C}^{2}}{\hbar^{2}}}
\end{equation}
Let us introduce the dimensionless variable $s=\frac{\left(p-p_{i}\right)r_{C}}{\hbar}$:
\begin{equation}
\frac{1}{\hbar^{3}}\int d\mathbf{p}\frac{1}{E_{p}^{\left(j\right)}}e^{-\frac{\left(\mathbf{p}-\mathbf{p}_{i}\right)^{2}r_{C}^{2}}{\hbar^{2}}}=\frac{\pi}{p_{i}r_{C}r_{c}^{3}}\int_{-\infty}^{+\infty}ds\frac{\left(s+p_{i}\frac{r_{C}}{\hbar}\right)}{\sqrt{\left(s+\frac{r_{C}}{\hbar}p_{i}\right)^{2}c^{2}+\frac{r_{C}^{2}}{\hbar^{2}}m_{j}^{2}c^{4}}}e^{-s^{2}}
\end{equation}
If $p_{i}c \gg \hbar c/r_{C} \sim 10\textrm{ eV}$ (the typical range of momenta of neutrinos is between $10^3\textrm{ eV}$ and $10^{19}\textrm{ eV}$) we can disregard $s$ both in the numerator and denominator, obtaining:

\begin{equation}
\frac{1}{\hbar^{3}}\int d\mathbf{p}\frac{1}{E_{p}^{\left(j\right)}}e^{-\frac{\left(\mathbf{p}-\mathbf{p}_{i}\right)^{2}r_{C}^{2}}{\hbar^{2}}}
\; \simeq \;
\frac{\pi}{r_{C}^{3}E_{i}^{\left(j\right)}}\int_{-\infty}^{+\infty}dse^{-s^{2}}
\; = \;
\frac{\pi^{3/2}}{r_{C}^{3}E_{i}^{\left(j\right)}},
\end{equation}
which is the desired result.

\section*{APPENDIX C: Approximation in the calculation of $I_{jk}^{\left(1\right)}$}

Here we justify the approximation we used to pass from Eq.~(\ref{eq:dasdasd}) to Eq.~(\ref{eq:syksdgar}). We start with Eq.~(\ref{eq:dasdasd}):
\begin{eqnarray} \label{eq:sdgyoudgs}
I_{jk}^{\left(1\right)}\left(\mathbf{p}_{i},s_{i};t\right)
& = &
\sum_{s_{f}}\int d\mathbf{p}_{f}\sqrt{\gamma_{m_{j}}\gamma_{m_{k}}}\frac{m_{j}m_{k}c^{4}}{\sqrt{E_{i}^{\left(j\right)}E_{f}^{\left(j\right)}E_{i}^{\left(k\right)}E_{f}^{\left(k\right)}}}\overline{u}(p_{f}^{\left(j\right)},s_{f})u(p_{i}^{\left(j\right)},s_{i})\overline{u}(p_{i}^{\left(k\right)},s_{i})u(p_{f}^{\left(k\right)},s_{f}) \nonumber \\
& & 
\times\frac{1-e^{\frac{i}{\hbar}\left(E_{f}^{\left(k\right)}-E_{i}^{\left(k\right)}-E_{f}^{\left(j\right)}+E_{i}^{\left(j\right)}\right)t}}{\frac{i}{\hbar}(E_{f}^{\left(j\right)}-E_{i}^{\left(j\right)}-E_{f}^{\left(k\right)}+E_{i}^{\left(k\right)})}\frac{1}{\left(2\pi\hbar\right)^{3}}e^{-\frac{\left(\mathbf{p}_{f}-\mathbf{p}_{i}\right)^{2}r_{C}^{2}}{\hbar^{2}}}
\end{eqnarray}
where:
\begin{equation} 
E_{f}^{\left(j\right)}=\sqrt{\mathbf{p}_{f}^{2}c^{2}+m_{j}^{2}c^{4}}\;\;\;\textrm{,}\;\;\; u\left(p,s\right)=\frac{p^{\mu}\gamma_{\mu}c+mc^{2}}{\sqrt{2mc^{2}\left(E_{p}+mc^{2}\right)}}u\left(0,s\right)\;\;\;\textrm{and}\;\;\; \sqrt{\gamma_{m_{j}}}=\sqrt{\gamma}\frac{m_{j}}{m_{0}}
\end{equation}
The first part of the integrand:
\begin{equation}
\frac{m_{j}m_{k}c^{4}}{\sqrt{E_{i}^{\left(j\right)}E_{f}^{\left(j\right)}E_{i}^{\left(k\right)}E_{f}^{\left(k\right)}}}\sum_{s_{f}}\overline{u}(p_{f}^{\left(j\right)},s_{f})u(p_{i}^{\left(j\right)},s_{i})\overline{u}(p_{i}^{\left(k\right)},s_{i})u(p_{f}^{\left(k\right)},s_{f})
\end{equation}
is a composition of polynomial functions of ${\bf p}_f$, and we can safely assume that it does not change too much, within the range where the Gaussian function is appreciably different from zero. Therefore we can then take  ${\bf p}_f = {\bf p}_i$; by using also the relation:
\begin{equation}
\overline{u}(p_{i}^{\left(1\right)},s_{f})u(p_{i}^{\left(1\right)},s_{i})=\delta_{s_{f},s_{i}},
\end{equation}
Eq.~(\ref{eq:sdgyoudgs}) becomes:
\begin{equation}
I_{jk}^{\left(1\right)}\left(\mathbf{p}_{i},s_{i};t\right)
\simeq
\sqrt{\gamma_{m_{j}}\gamma_{m_{k}}}\frac{m_{j}m_{k}c^{4}}{E_{i}^{\left(j\right)}E_{i}^{\left(k\right)}}\frac{1}{\left(2\pi\hbar\right)^{3}}\underset{\equiv I}{\underbrace{\int d\mathbf{p}_{f}\frac{1-e^{\frac{i}{\hbar}(E_{f}^{\left(k\right)}-E_{i}^{\left(k\right)}-E_{f}^{\left(j\right)}+E_{i}^{\left(j\right)})t}}{\frac{i}{\hbar}(E_{f}^{\left(j\right)}-E_{i}^{\left(j\right)}-E_{f}^{\left(k\right)}+E_{i}^{\left(k\right)})}e^{-\frac{(\mathbf{p}_{f}-\mathbf{p}_{i})^{2}r_{C}^{2}}{\hbar^{2}}}}}
\end{equation}
Now we have to focus our attention on the integral $I$, which contains an oscillating term that needs special care. As before, we write the integral in polar coordinates and perform the integration over the angular variables; moreover, we introduce once again the a-dimensional variable $s=\frac{\left(p_{f}-p_{i}\right)r_{C}}{\hbar}$. We have:
\begin{equation} \label{eq:iufsw}
I=\pi\frac{\hbar^{3}}{p_{i}r_{C}^{3}}t\int_{-\infty}^{+\infty}ds\left(\frac{\hbar}{r_{C}}s+p_{i}\right)\frac{e^{ig\left(s\right)}-1}{ig\left(s\right)}e^{-s^{2}},
\end{equation}
where we have defined:
\begin{eqnarray} \label{eq:sdsdtukz}
g\left(s\right)
& \equiv &
\frac{1}{\hbar}(E_{f}^{\left(k\right)}-E_{i}^{\left(k\right)}-E_{f}^{\left(j\right)}+E_{i}^{\left(j\right)})t \nonumber \\
& = &
\frac{ct}{r_{C}}\left(\sqrt{\left(s+y\right)^{2}+a_{k}}-\sqrt{\left(s+y\right)^{2}+a_{j}}-\sqrt{y^{2}+a_{k}}+\sqrt{y^{2}+a_{j}}\right),
\end{eqnarray}
with
\begin{equation}
a_{j}\equiv\frac{r_{C}^{2}}{\hbar^{2}}m_{j}^{2}c^{2}=\left(10^{-2}\textrm{eV}^{-2}\right)m_{j}^{2}c^{4}\;\;\;\textrm{e}\;\;\; y\equiv\frac{r_{C}}{\hbar}p_{i}=\left(10^{-1}\textrm{eV}^{-1}\right)p_{i}c
\end{equation}
Our goal is to show that $g\left(s\right)$ does not vary appreciably, within the range where the Gaussian term is significantly different from zero, and can be approximated with $g\left(0\right)=0$; in this way, the integral can be computed exactly. This kind of approximation is not obvious because the factor $ct/r_{C}$ in front of Eq.~(\ref{eq:sdsdtukz}) can be very big.

In the ultra-relativistic limit, we approximate the particle's velocity with the speed of light. In this limit, $y^{2} \gg a_{j},a_{k}$, and so we can expand the square roots in $g(s)$ using the Taylor series $\sqrt{x+\epsilon}=\sqrt{x}+\frac{\epsilon}{2\sqrt{x}}$ and we get:
\begin{equation}
g\left(s\right)\simeq\frac{ct}{r_{C}}\left(a_{j}-a_{k}\right)\frac{s}{\left(s+y\right)y}=\frac{ct}{r_{C}}\frac{\left(a_{j}-a_{k}\right)}{\left(y+\frac{y^{2}}{s}\right)}.
\end{equation}
In order for $g(s)$ to remain small within the interval where the Gaussian term of Eq.~(\ref{eq:iufsw}) is appreciably different from zero, we need:
\begin{equation}\label{eq:condizione}
\left(y+\frac{y^{2}}{s}\right) \; \gg \; \frac{ct}{r_{C}}\left(a_{j}-a_{k}\right)
\end{equation}
In all physically interesting situations, the term on the right hand side of Eq.~(\ref{eq:condizione}) is bigger than 1; moreover $s$ is of the order of unity, because of the Gaussian in Eq.~(\ref{eq:iufsw}). Inequality~\eqref{eq:condizione} is verified if the following condition is true:
\begin{equation}
y \; \gg \; \sqrt{\frac{ct}{r_{C}}\left(a_{j}-a_{k}\right)}.
\end{equation}
Typically, cosmogenic neutrinos have energies bigger than $10^{18}\textrm{eV}$ and travel distances of at most $10^{9}$ light-years~\cite{Ch}. This means that, in the worst case, $ct/r_{C} \sim 10^{32}$ while $(a_{j}-a_{k})$ $ =(10^{-2}\textrm{eV}^{-2})$ $(m_{j}^{2}c^{4}-m_{k}^{2}c^{4})\simeq(10^{-2}\textrm{eV}^{-2})(2\times10^{-3}\textrm{eV}^{2})=10^{-5}$. So we must have $y \gg 10^{14}$, that means $p_{i}c = y/(10^{-1}\textrm{eV}^{-1}) \gg 10^{15}\textrm{eV}$, which is satisfied.

For atmospheric neutrinos, $ct/r_{C}$ is in the range $10^{11}-10^{14}$ while the range of energies is between $10^{-1}\textrm{GeV}$ and $10^{4}\textrm{GeV}$~\cite{Neu1}. This means that, even in the worse case, the condition to check is $y \gg 10^{5}$, which means $p_{i}c = y/(10^{-1}\textrm{eV}^{-1}) \gg 10^{6}\textrm{eV}$. This is also satisfied.

\end{document}